\documentclass[a4paper,fleqn]{article}
\usepackage{amsmath}
\usepackage[dvips]{graphicx}
\begin{document}

\title{\textbf{Integrability of a generalized short pulse equation revisited}}

\author{\textsc{Sergei Sakovich}\bigskip \\
\small Institute of Physics, National Academy of Sciences of Belarus \\
\small sergsako@gmail.com}

\date{}

\maketitle

\begin{abstract}
We further generalize the generalized short pulse equation studied recently in [Commun. Nonlinear Sci. Numer. Simulat. 39 (2016) 21--28; arXiv:1510.08822], and find in this way two new integrable nonlinear wave equations which are transformable to linear Klein--Gordon equations.
\end{abstract}

\section{Introduction}

In this paper, we study the integrability of the nonlinear wave equation
\begin{equation}
u_{xt} = a u^2 u_{xx} + b u u_x^2 \label{e1}
\end{equation}
containing two arbitrary parameters, $a$ and $b$, not equal zero simultaneously. Actually, there is only one essential parameter in \eqref{e1}, the ratio $a/b$ or $b/a$, which is invariant under the scale transformations of $u$, $x$ and $t$, while the values of $a$ and $b$ are not invariant. We show that this equation \eqref{e1} is integrable in two (and, most probably, only two) distinct cases, namely, when $a/b = 1/2$ and $a/b = 1$, which correspond via scale transformations of variables to the equations
\begin{equation}
u_{xt} = \frac{1}{6} \left( u^3 \right)_{xx} \label{e2}
\end{equation}
and
\begin{equation}
u_{xt} = \frac{1}{2} u \left( u^2 \right)_{xx} , \label{e3}
\end{equation}
respectively.

There is the following reason to study the nonlinear equation \eqref{e1}. Recently, in \cite{S1}, we studied the integrability of the generalized short pulse equation
\begin{equation}
u_{xt} = u + a u^2 u_{xx} + b u u_x^2 \label{e4}
\end{equation}
containing two arbitrary parameters, $a$ and $b$, not equal zero simultaneously. We showed that there are two (and, most probably, only two) integrable cases of \eqref{e4}, namely, those with $a/b = 1/2$ and $a/b = 1$, which can be written as
\begin{equation}
u_{xt} = u + \frac{1}{6} \left( u^3 \right)_{xx} \label{e5}
\end{equation}
and
\begin{equation}
u_{xt} = u + \frac{1}{2} u \left( u^2 \right)_{xx} \label{e6}
\end{equation}
via scale transformations of variables. The nonlinear equation \eqref{e5} is the celebrated short pulse equation which appeared first in differential geometry \cite{BRT,R}, was later rediscovered in nonlinear optics \cite{SW,CJSW}, and since then has been studied in almost any aspect of its integrability \cite{SS1,B1,B2,SS2,SS3,M1,M2,P1,LPS,PS,P2,FMO}. The nonlinear equation \eqref{e6}, called the single-cycle pulse equation (due to the property of its smooth envelope soliton solution) \cite{S1} or the modified short pulse equation \cite{M3}, is a scalar reduction of the integrable system of coupled short pulse equations of Feng \cite{F}. One may wonder, looking at \eqref{e4}, why not to generalize this equation further, as
\begin{equation}
u_{xt} = a u^2 u_{xx} + b u u_x^2 + c u \label{e7}
\end{equation}
with arbitrary parameters $a$, $b$ and $c$, in order to find new integrable nonlinear wave equations in this way. It is easy to see, however, that there are only two essentially different values of the parameter $c$ in \eqref{e7}, namely, $c=0$ and (without loss of generality) $c=1$, because one can always make $c=1$ by a scale transformation of variables if $c \ne 0$. Since the case of \eqref{e7} with $c=1$ is the nonlinear equation \eqref{e4} studied in \cite{S1}, we concentrate in the present paper on the remaining case of \eqref{e7} with $c=0$, which is the nonlinear equation \eqref{e1}.

In Section~\ref{s2} of this paper, we transform the nonlinear equation \eqref{e1} with any finite value of $a/b$ to a corresponding (in general, nonlinear) Klein--Gordon equation whose nonlinearity depends on $a/b$, and we bring \eqref{e1} with $b=0$ into a form suitable for the Painlev\'{e} analysis. In Section~\ref{s3}, using the known results on integrability of nonlinear Klein--Gordon equations (for $b \ne 0$) and the Painlev\'{e} test for integrability (for $b=0$), we show that the nonlinear equation \eqref{e1} is integrable if (and, most probably, only if) $a/b = 1/2$ or $a/b = 1$, that is, when the nonlinear equation \eqref{e1} is transformable to linear Klein--Gordon equations. This allows us to obtain parametric representations for general solutions of the nonlinear equations \eqref{e2} and \eqref{e3} and discuss their properties. Section~\ref{s4} contains concluding remarks.

\section{Transformation} \label{s2}

In our experience, a transformation found to relate a new nonlinear equation with a known old one is a powerful tool to derive the fact and character of integrability or non-integrability of the new equation from what is known on integrability or non-integrability of the old equation \cite{S1,S2,S3,S4,S5}. By means of transformations relating new equations with known old ones, it is possible to derive analytic properties of solutions \cite{S6}, expressions for special and general solutions \cite{S1,SS2,SS3,S7,S8}, Lax pairs, Hamiltonian structures and recursion operators \cite{S9,BS1,BS2,BS3} of the new equations from the corresponding known properties and objects of the old equations.

When $a=0$ in \eqref{e1}, we have $b \ne 0$, and we make $b=1$ by a scale transformation of variables, without loss of generality,
\begin{equation}
u_{xt} = u u_x^2 . \label{e8}
\end{equation}
If $u_x \ne 0$, we rewrite \eqref{e8} as
\begin{equation}
\left( \frac{1}{u_x} \right)_t + u = 0 , \label{e9}
\end{equation}
introduce the new dependent variable $v(x,t)$,
\begin{equation}
v = \frac{1}{u_x} , \label{e10}
\end{equation}
and get the nonlinear Klein--Gordon equation
\begin{equation}
v_{xt} = - \frac{1}{v} . \label{e11}
\end{equation}
The inverse transformation from \eqref{e11} to \eqref{e8},
\begin{equation}
u = - v_t , \label{e12}
\end{equation}
is also a local transformation, that is, like \eqref{e10}, it requires no integration. Note that the transformations \eqref{e10} and \eqref{e12} between the equations \eqref{e8} and \eqref{e11} do not cover the case of $u_x = 0$. However, $u=u(t)$ with any function $u(t)$ satisfies the nonlinear equation \eqref{e1} with any values of $a$ and $b$, and this set of special solutions tells nothing about the integrability of \eqref{e8}.

When $a \ne 0$ in \eqref{e1}, we introduce the new independent variable $y$,
\begin{equation}
x = x(y,t) , \qquad u(x,t) = p(y,t) , \label{e13}
\end{equation}
and impose the condition
\begin{equation}
x_t = - a p^2 \label{e14}
\end{equation}
on the function $x(y,t)$ to considerably simplify the result. Then the studied equation \eqref{e1} takes the form
\begin{equation}
x_y p_{yt} + (2a-b) p p_y^2 = 0 . \label{e15}
\end{equation}
This equation \eqref{e15} is invariant under the transformation $y \mapsto Y(y)$ with any function $Y$, which means that solutions of the system \eqref{e14} and \eqref{e15} determine solutions $u(x,t)$ of \eqref{e1} parametrically, with $y$ being the parameter. Next, we make use of the new dependent variable $q(y,t)$, such that
\begin{equation}
x_y = \frac{1}{q} p_y , \label{e16}
\end{equation}
which means that $q(y,t) = u_x (x,t)$. Since $q \ne 0$ in \eqref{e16}, our transformation does not cover the evident special solutions of \eqref{e1} with $u_x = 0$. The compatibility condition $( x_t )_y = ( x_y )_t$ for \eqref{e14} and \eqref{e16} reads
\begin{equation}
p_{yt} = \frac{1}{q} p_y q_t - 2a p q p_y . \label{e17}
\end{equation}
Eliminating $x_y$ from \eqref{e15} and \eqref{e16}, and using \eqref{e17}, we get
\begin{equation}
q_t = b p q^2 . \label{e18}
\end{equation}
Due to \eqref{e18}, we have to consider the cases of $b \ne 0$ and $b=0$ separately.

If $b \ne 0$, we make $b=1$ by a scale transformation of variables, without loss of generality. Using the new dependent variable $r(y,t)$,
\begin{equation}
r = \frac{1}{q} , \label{e19}
\end{equation}
we get
\begin{equation}
p = - r_t \label{e20}
\end{equation}
from \eqref{e18}, and
\begin{equation}
r_{ytt} = \frac{2a-1}{r} r_t r_{yt} \label{e21}
\end{equation}
from \eqref{e17}. Dividing the left- and right-hand sides of \eqref{e21} by $r_{yt}$ ($r_{yt} \ne 0$ if $x_y \ne 0$, owing to \eqref{e20} and \eqref{e16}), and integrating over $t$, we get
\begin{equation}
r_{yt} = h(y) r^{2a-1} \label{e22}
\end{equation}
with any nonzero function $h(y)$, which appeared as the (exponent of) ``constant'' of integration. Finally, we make $h(y)=1$ in \eqref{e22} by the transformation $y \mapsto Y(y)$ with a properly chosen $Y(y)$ (thus suppressing the arbitrariness of the parameter $y$ down to $y \mapsto y + \mathrm{constant}$), and obtain the following result. All solutions of the considered case of \eqref{e1},
\begin{equation}
u_{xt} = a u^2 u_{xx} + u u_x^2 , \label{e23}
\end{equation}
except for solutions with $u_x = 0$, are determined parametrically by solutions of the nonlinear Klein--Gordon equation
\begin{equation}
r_{yt} = r^{2a-1} \label{e24}
\end{equation}
via the relations
\begin{gather}
u(x,t) = - r_t (y,t) , \notag \\
x = x(y,t) : \qquad x_y = - r^{2a} , \qquad x_t = - a r_t^2 , \label{e25}
\end{gather}
where $y$ serves as the parameter, and $a$ is an arbitrary nonzero constant.

If $b=0$, we have $a \ne 0$, and we make $a=1$ by a scale transformation of variables, without loss of generality. In this case, we get $q_t = 0$ from \eqref{e18}, that is, $q=q(y)$ with any nonzero function $q(y)$, and the equation \eqref{e17} takes the form
\begin{equation}
p_{yt} + 2 q(y) p p_y = 0 . \label{e26}
\end{equation}
Consequently, all solutions of the considered case of \eqref{e1},
\begin{equation}
u_{xt} = u^2 u_{xx} , \label{e27}
\end{equation}
except for solutions with $u_x = 0$, are determined parametrically by solutions of the nonlinear equation \eqref{e26} with any $q(y) \ne 0$ via the relations
\begin{gather}
u(x,t) = p(y,t) , \notag \\
x = x(y,t) : \qquad x_y = \frac{1}{q(y)} p_y , \qquad x_t = - p^2 , \label{e28}
\end{gather}
where $y$ serves as the parameter. Note that the arbitrariness of $q(y)$ in \eqref{e26} cannot be suppressed by the change of parametrization $y \mapsto Y(y)$.

\section{Integrability} \label{s3}

Integrability of nonlinear Klein--Gordon equations is very well studied. It was shown in \cite{ZS} that the equation
\begin{equation}
z_{\xi \eta} = w(z) \label{e29}
\end{equation}
possesses a nontrivial group of higher symmetries if and only if the function $w(z)$ satisfies either the condition
\begin{equation}
w' = \alpha w \label{e30}
\end{equation}
or the condition
\begin{equation}
w'' = \alpha w + \beta w' , \label{e31}
\end{equation}
where $z = z( \xi , \eta )$, the prime denotes the derivative with respect to $z$, the constant $\alpha$ in \eqref{e30} is arbitrary, while the constants $\alpha$ and $\beta$ in \eqref{e31} must satisfy the condition
\begin{equation}
\beta \left( \alpha - 2 \beta^2 \right) = 0 . \label{e32}
\end{equation}
No more integrable cases of \eqref{e29} have been discovered by various methods as yet.

The right-hand side of the nonlinear Klein--Gordon equation \eqref{e11} satisfies neither \eqref{e30} nor \eqref{e31}. Therefore this equation, together with the corresponding case \eqref{e8} of the studied equation \eqref{e1}, must be non-integrable. The right-hand side of the nonlinear Klein--Gordon equation \eqref{e24} satisfies \eqref{e30} or \eqref{e31} for two values of $a$ only, $a = 1/2$ or $a=1$, when \eqref{e24} is actually a linear equation, while the corresponding nonlinear equation \eqref{e23} takes the form \eqref{e2} or \eqref{e3}, respectively.

The case of \eqref{e24} with $a = 1/2$ is the Darboux integrable linear equation
\begin{equation}
r_{yt} = 1 \label{e33}
\end{equation}
whose solutions parametrically determine all solutions (except for solutions with $u_x = 0$) of the nonlinear equation \eqref{e2} via the relations
\begin{gather}
u(x,t) = - r_t (y,t) , \notag \\
x = x(y,t) : \qquad x_y = - r , \qquad x_t = - \frac{1}{2} r_t^2 , \label{e34}
\end{gather}
where $y$ serves as the parameter. Taking the general solution of \eqref{e33}
\begin{equation}
r = y t + f(y) + g(t) , \label{e35}
\end{equation}
where $f(y)$ and $g(t)$ are arbitrary functions, we obtain via \eqref{e34} the following parametric representation for the general solution of the nonlinear equation \eqref{e2}:
\begin{gather}
u(x,t) = - y - g'(t) , \notag \\
x = - \frac{1}{2} y^2 t - \int{f(y) \, dy} - y g(t) - \frac{1}{2} \int{[ g'(t) ]^2 \, dt} , \label{e36}
\end{gather}
where the prime stands for the derivative. It follows from \eqref{e36} that
\begin{equation}
u_x = \frac{1}{y t + f(y) + g(t)} , \label{e37}
\end{equation}
which shows that the general solution \eqref{e36} does not cover the evident special solutions of \eqref{e2} with $u_x = 0$. Also, due to \eqref{e37}, there are apparently no solutions of \eqref{e2} without singularities of the type $u_x \to \infty$, besides the solutions with $u_x = 0$. We do not see how to choose the functions $f(y)$ and $g(t)$ to make the denominator in \eqref{e37} not equal zero for all values of $y$ and $t$.

\begin{figure}
\includegraphics[width=12cm]{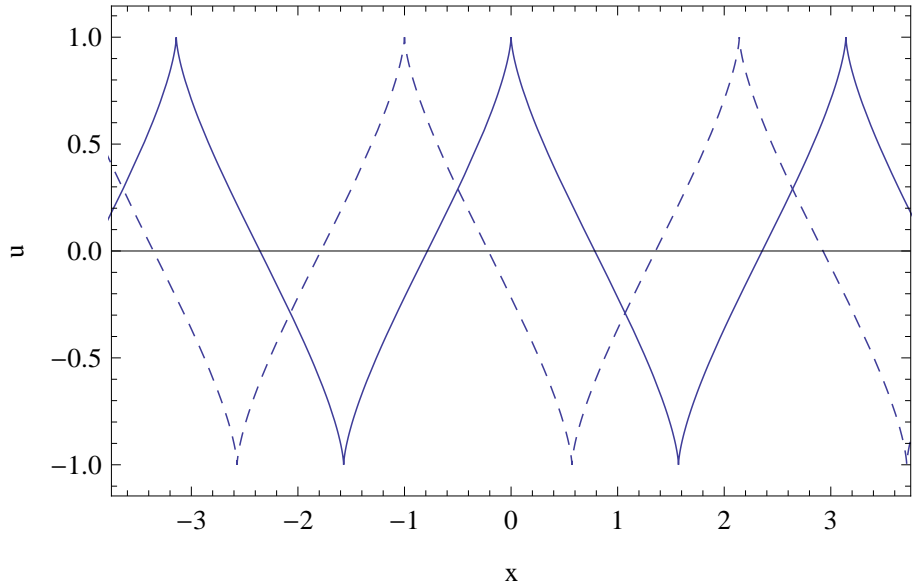}
\caption{The singular solution \eqref{e42}: $t = 0$ (solid) and $t = 1$ (dashed). \label{f1}}
\end{figure}

\begin{figure}
\includegraphics[width=12cm]{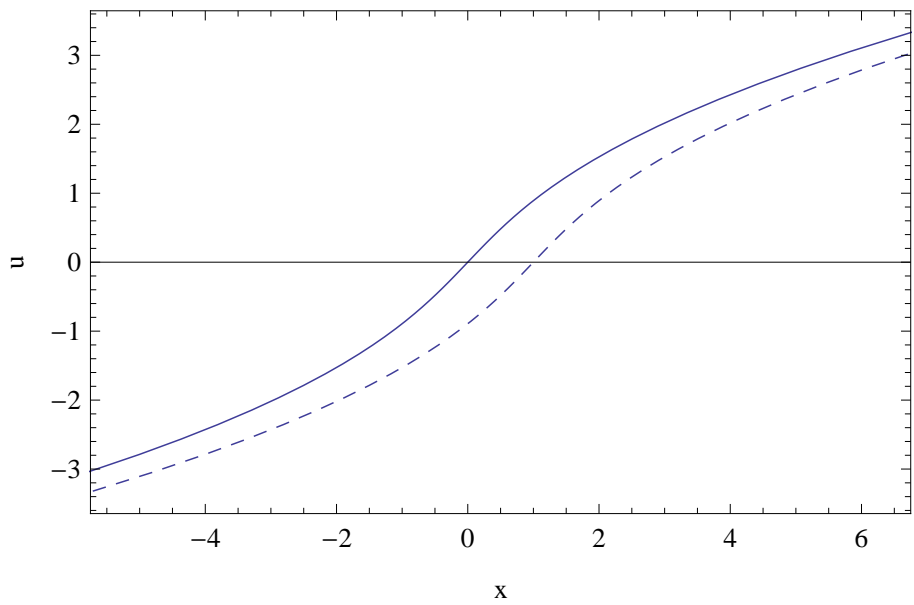}
\caption{The smooth solution \eqref{e44}: $t = 0$ (solid) and $t = 1$ (dashed). \label{f2}}
\end{figure}

The case of \eqref{e24} with $a=1$ is the Fourier integrable linear equation
\begin{equation}
r_{yt} = r \label{e38}
\end{equation}
whose solutions parametrically determine all solutions (except for solutions with $u_x = 0$) of the nonlinear equation \eqref{e3} via the relations
\begin{gather}
u(x,t) = - r_t (y,t) , \notag \\
x = x(y,t) : \qquad x_y = - r^2 , \qquad x_t = - r_t^2 , \label{e39}
\end{gather}
where $y$ serves as the parameter. Since
\begin{equation}
u_x = \frac{1}{r} \label{e40}
\end{equation}
due to \eqref{e39}, the parametric representation \eqref{e39} of the general solution of \eqref{e3} does not cover the evident special solutions of \eqref{e3} with $u_x = 0$. It is easy to see from \eqref{e40} that a solution $u(x,t)$ of \eqref{e3} contains singularities of the type $u_x \to \infty$ if the corresponding solution $r(y,t)$ of \eqref{e38} contains zeroes. For example, if we take
\begin{equation}
r = \sin (y-t) , \label{e41}
\end{equation}
we get from \eqref{e39} the solution
\begin{equation}
u = \cos (y-t) , \qquad x = - \frac{1}{2} (y+t) + \frac{1}{4} \sin 2(y-t) \label{e42}
\end{equation}
containing singularities, as shown in Figure~\ref{f1}.
On the contrary, taking
\begin{equation}
r = \cosh (y+t) , \label{e43}
\end{equation}
we get the smooth solution
\begin{equation}
u = - \sinh (y+t) , \qquad x = - \frac{1}{2} (y-t) - \frac{1}{4} \sinh 2(y+t) , \label{e44}
\end{equation}
shown in Figure~\ref{f2}.
Note that, in \eqref{e42} and \eqref{e44}, the constant of integration in $x$ has been fixed so that $x|_{y=t=0} = 0$.

It only remains to test the integrability of the nonlinear equation \eqref{e27}, because the case of \eqref{e1} with $b=0$ could not be transformed into a Klein--Gordon equation. We have found the transformation \eqref{e28} which relates \eqref{e27} with the nonlinear equation \eqref{e26}. Let us study the integrability of \eqref{e26} by means of the Painlev\'{e} analysis \cite{WTC,T,H}, which is, in our experience, a reliable and easy-to-use tool to test the integrability of nonlinear equations \cite{S10,S11,S12,S13,S14,KS1,S15,S16,KSY,KS2,S17,S18,S19,S20,S21,S22}. The reliability of the Painlev\'{e} test for integrability has been empirically verified in numerous studies of multi-parameter families of nonlinear equations, including the fifth-order KdV-type equation \cite{HO}, the coupled KdV equations \cite{K,S23,S24,KKS1,S25}, the symmetrically coupled higher-order nonlinear Schr\"{o}dinger equations \cite{ST1,ST2,ST3}, the generalized Ito system \cite{KKS2}, the sixth-order bidirectional wave equation \cite{KKSST}, and the seventh-order KdV-type equation \cite{X}.

A hypersurface $\phi(y,t) = 0$ is non-characteristic for the studied equation \eqref{e26} if $\phi_y \phi_t \ne 0$, and we choose $\phi_t = 1$ without loss of generality, that is, $\phi = t + \psi(y)$ with $\psi_y \ne 0$. Using the expansion
\begin{equation}
p = p_0 (y) \phi^\gamma + \dotsb +  p_n (y) \phi^{\gamma + n} + \dotsb , \label{e45}
\end{equation}
we find the dominant singular behavior of solutions of \eqref{e26} near $\phi = 0$,
\begin{equation}
\gamma = -1 , \qquad p_0 = \frac{1}{q(y)} , \label{e46}
\end{equation}
together with the corresponding positions of resonances,
\begin{equation}
n = -1 , 2 , \label{e47}
\end{equation}
where $n = -1$ refers to the arbitrariness of $\psi(y)$. Substituting the expansion
\begin{equation}
p = p_0 (y) \phi^{-1} + p_1 (y) + p_2 (y) \phi + \dotsb \label{e48}
\end{equation}
to \eqref{e26}, and collecting terms with equal degrees of $\phi$, we get the following. The terms with $\phi^{-3}$, of course, give the expression \eqref{e46} for $p_0$. The terms with $\phi^{-2}$ give the expression
\begin{equation}
p_1 = \frac{- q_y}{2 q^2 \psi_y} . \label{e49}
\end{equation}
The terms with $\phi^{-1}$, however, do not determine $p_2 (y)$ (here we have the resonance) but lead to the nontrivial compatibility condition
\begin{equation}
q q_y \psi_{yy} -q q_{yy} \psi_y + 3 q_y^2 \psi_y = 0 . \label{e50}
\end{equation}
In order to satisfy this condition \eqref{e50} for all functions $\psi(y)$ ($\psi_y \ne 0$), we must set $q_y = 0$. Otherwise, for $q_y \ne 0$, the compatibility condition \eqref{e50} is not satisfied identically, and we must introduce logarithmic terms to the expansion \eqref{e48}, starting from the term proportional to $\phi \log \phi$, which is a clear indication of non-integrability. Consequently, the nonlinear equation \eqref{e26} is integrable for $q = \mathrm{constant}$ only, not for any nonzero function $q(y)$. Therefore the corresponding equation \eqref{e27} is not integrable. Moreover, since $q(y,t) = u_x (x,t)$, we believe that the only solutions of the nonlinear equation \eqref{e27} obtainable in a closed form are the evident solutions with $u_x = \mathrm{constant}$.

\section{Conclusion} \label{s4}

In this paper, we have generalized further the generalized short pulse equation studied recently in \cite{S1}, and found in this way two new integrable nonlinear wave equations, namely, \eqref{e2} and \eqref{e3}, which are transformable to linear Klein--Gordon equations. These new equations \eqref{e2} and \eqref{e3}, due to the absence of the linear term ``$u$'' in them, can be considered as ``massless'' counterparts of the short pulse equation \eqref{e5} and the single-cycle pulse equation \eqref{e6}, respectively. Let us note, however, that the types of integrability of \eqref{e2} and \eqref{e3} are essentially different from the type of integrability of \eqref{e5} and \eqref{e6}. While the equations \eqref{e5} and \eqref{e6} are two ``avatars'' (in the sense of transformations) of the sine-Gordon equation, the new nonlinear equation \eqref{e2} is an ``avatar'' of a Darboux integrable linear Klein--Gordon equation, and the new nonlinear equation \eqref{e3} is an ``avatar'' of a Fourier integrable linear Klein--Gordon equation. Taking this into account, we expect that the integrability properties of the new equation \eqref{e2} are similar to those of the Liouville equation (continual sets of generalized symmetries and conservation laws, and several mutually non-equivalent Lax pairs \cite{S26}), whereas the properties of \eqref{e3} may be similar to those of linear wave equations (a discrete hierarchy of symmetries, a finite set of conservation laws, and no phase shifts in wave interactions). We believe that these new equations \eqref{e2} and \eqref{e3} can be useful, as integrable scalar reductions, for classifications of integrable vector short pulse equations.

Let us also note that our equations \eqref{e2} and \eqref{e3} did not appear in the most recent integrability classification of generalized short pulse equations of Hone, Novikov and Wang \cite{HNW} because equations without the linear term ``$u$'' were not studied there.


\begin{thebibliography}{99}

\small

\bibitem{S1} S. Sakovich, Transformation and integrability of a generalized short pulse equation, Commun. Nonlinear Sci. Numer. Simulat. 39 (2016) 21--28; arXiv: 1510.08822.

\bibitem{BRT} R. Beals, M. Rabelo, K. Tenenblat, B\"{a}cklund transformations and inverse scattering solutions for some pseudospherical surface equations, Stud. Appl. Math. 81 (1989) 125--151.

\bibitem{R} M.L. Rabelo, On equations which describe pseudospherical surfaces, Stud. Appl. Math. 81 (1989) 221--248.

\bibitem{SW} T. Sch\"{a}fer, C.E. Wayne, Propagation of ultra-short optical pulses in cubic nonlinear media, Physica D 196 (2004) 90--105.

\bibitem{CJSW} Y. Chung, C.K.R.T. Jones, T. Sch\"{a}fer, C.E. Wayne, Ultra-short pulses in linear and nonlinear media, Nonlinearity 18 (2005) 1351--1374; arXiv: nlin/0408020.

\bibitem{SS1} A. Sakovich, S. Sakovich, The short pulse equation is integrable, J. Phys. Soc. Jpn. 74 (2005) 239--241; arXiv:nlin/0409034.

\bibitem{B1} J.C. Brunelli, The short pulse hierarchy, J. Math. Phys. 46 (2005) 123507; arXiv:nlin/0601015.

\bibitem{B2} J.C. Brunelli, The bi-Hamiltonian structure of the short pulse equation, Phys. Lett. A 353 (2006) 475--478; arXiv:nlin/0601014.

\bibitem{SS2} A. Sakovich, S. Sakovich, Solitary wave solutions of the short pulse equation, J. Phys. A 39 (2006) L361--L367; arXiv:nlin/0601019.

\bibitem{SS3} A. Sakovich, S. Sakovich, On transformations of the Rabelo equations, SIGMA 3 (2007) 086; arXiv:0705.2889.

\bibitem{M1} Y. Matsuno, Multiloop soliton and multibreather solutions of the short pulse model equation, J. Phys. Soc. Jpn. 76 (2007) 084003.

\bibitem{M2} Y. Matsuno, Periodic solutions of the short pulse model equation, J. Math. Phys. 49 (2008) 073508.

\bibitem{P1} E.J. Parkes, Some periodic and solitary travelling-wave solutions of the short-pulse equation, Chaos Solitons Fract. 38 (2008) 154--159.

\bibitem{LPS} Y. Liu, D. Pelinovsky, A. Sakovich, Wave breaking in the short-pulse equation, Dyn. Part. Diff. Eqs. 6 (2009) 291--310; arXiv:0905.4668.

\bibitem{PS} D. Pelinovsky, A. Sakovich, Global well-posedness of the short-pulse and sine-Gordon equations in energy space, Commun. Part. Diff. Eqs. 35 (2010)  613--629; arXiv:0809.5052.

\bibitem{P2} E.J. Parkes, A note on loop-soliton solutions of the short-pulse equation, Phys. Lett. A 374 (2010) 4321--4323.

\bibitem{FMO} B.F. Feng, K. Maruno, Y. Ohta, Integrable discretizations of the short pulse equation, J. Phys. A 43 (2010) 085203; arXiv:0912.1914.

\bibitem{M3} Y. Matsuno, Integrable multi-component generalization of a modified short pulse equation, J. Math. Phys. 57 (2016) 111507; arXiv:1607.01079.

\bibitem{F} B.F. Feng, An integrable coupled short pulse equation, J. Phys. A 45 (2012) 085202.

\bibitem{S2} S.Yu. Sakovich, Fujimoto--Watanabe equations and differential substitutions, J. Phys. A 24 (1991) L519--L521.

\bibitem{S3} S.Yu. Sakovich, On Miura transformations of evolution equations, J. Phys. A 26 (1993) L369--L373.

\bibitem{S4} S.Yu. Sakovich, Transformation of a generalized Harry Dym equation into the Hirota--Satsuma system, Phys. Lett. A 321 (2004) 252--254; arXiv:nlin/0309077.

\bibitem{S5} S.Yu. Sakovich, On bosonic limits of two recent supersymmetric extensions of the Harry Dym hierarchy, J. Math. Phys. 45 (2004) 2338--2342; arXiv:nlin/0310039.

\bibitem{S6} S.Yu. Sakovich, The Painlev\'{e} property transformed, J. Phys. A 25 (1992) L833--L836.

\bibitem{S7} S. Sakovich, On a Whitham-type equation, SIGMA 5 (2009) 101; arXiv: 0909.4455.

\bibitem{S8} S. Sakovich, Smooth soliton solutions of a new integrable equation by Qiao, J. Math. Phys. 52 (2011) 023509; arXiv:1010.1907.

\bibitem{S9} S.Yu. Sakovich, On integrability of one third-order nonlinear evolution equation, Phys. Lett. A 314 (2003) 232--238; arXiv:nlin/0303040.

\bibitem{BS1} J.C. Brunelli, S. Sakovich, On integrability of the Yao--Zeng two-component short-pulse equation, Phys. Lett. A 377 (2012) 80--82; arXiv:1205.6969.

\bibitem{BS2} J.C. Brunelli, S. Sakovich, Hamiltonian structures for the Ostrovsky--Vakh\-nenko equation, Commun. Nonlinear Sci. Numer. Simulat. 18 (2013) 56--62; arXiv:1202.5129.

\bibitem{BS3} J.C. Brunelli, S. Sakovich, Hamiltonian integrability of two-component short pulse equations, J. Math. Phys. 54 (2013) 012701; arXiv:1210.5265.

\bibitem{ZS} A.V. Zhiber, A.B. Shabat, Klein--Gordon equations with a nontrivial group, Sov. Phys. Dokl. 24 (1979) 607--609.

\bibitem{WTC} J. Weiss, M. Tabor, G. Carnevale, The Painlev\'{e} property  for partial differential equations, J. Math. Phys. 24 (1983) 522--526.

\bibitem{T} M. Tabor, Chaos and Integrability in Nonlinear Dynamics: An Introduction, Wiley, New York, 1989.

\bibitem{H} A.N.W. Hone, Painlev\'{e} tests, singularity structure and integrability, In: A.V. Mikhailov (Ed.), Integrability, Lect. Notes in Phys. 767, Springer, Berlin, 2009, 245--277; arXiv:nlin.SI/0502017.

\bibitem{S10} S.Yu. Sakovich, Painlev\'{e} analysis of new soliton equations by Hu, J. Phys. A 27 (1994) L503--L505.

\bibitem{S11} S.Yu. Sakovich, Painlev\'{e} analysis and B\"{a}cklund transformations of Doktorov--Vlasov equations, J. Phys. A 27 (1994) L33--L38.

\bibitem{S12} S.Yu. Sakovich, On zero-curvature representations of evolution equations, J. Phys. A 28 (1995) 2861--2869.

\bibitem{S13} S.Yu. Sakovich, Painlev\'{e} analysis of a higher-order nonlinear Schr\"{o}dinger equation, J. Phys. Soc. Jpn. 66 (1997) 2527--2529.

\bibitem{S14} S.Yu. Sakovich, On integrability of a $(2+1)$-dimensional perturbed KdV equation, J. Nonlinear Math. Phys. 5 (1998) 230--233; arXiv:solv-int/9805012.

\bibitem{KS1} A. Karasu-Kalkanl\i, S.Yu. Sakovich, B\"{a}cklund transformation and special solutions for the Drinfeld--Sokolov--Satsuma--Hirota system of coupled equations, J. Phys. A 34 (2001) 7355--7358; arXiv:nlin/0102001.

\bibitem{S15} S.Yu. Sakovich, A system of four ODEs: the singularity analysis, J. Nonlinear Math. Phys. 8 (2001) 217--219; arXiv:nlin/0003039.

\bibitem{S16} S.Yu. Sakovich, On integrability of differential constraints arising from the singularity analysis, J. Nonlinear Math. Phys. 9 (2002) 21--25; arXiv: nlin/0004037.

\bibitem{KSY} A. Karasu-Kalkanl\i, S.Yu. Sakovich, \'{I}. Yurdu\c{s}en, Integrability of Kersten--Krasil'shchik coupled KdV--mKdV equations: singularity analysis and Lax pair, J. Math. Phys. 44 (2003) 1703--1708; arXiv:nlin/0206046.

\bibitem{KS2} A. Karasu-Kalkanl\i, S. Sakovich, Singularity analysis of a spherical Kadomtsev--Petviashvili equation, J. Phys. Soc. Jpn. 74 (2005) 505--507; arXiv:nlin/0404037.

\bibitem{S17} S. Sakovich, Enlarged spectral problems and nonintegrability, Phys. Lett. A 345 (2005) 63--68; arXiv:nlin/0504037.

\bibitem{S18} S. Sakovich, On a ``mysterious'' case of a quadratic Hamiltonian, SIGMA 2 (2006) 064; arXiv:nlin/0408027.

\bibitem{S19} S. Sakovich, Integrability of the vector short pulse equation, J. Phys. Soc. Jpn. 77 (2008) 123001; arXiv:0801.3179.

\bibitem{S20} S. Sakovich, Singularity analysis and integrability of a Burgers-type system of Foursov, SIGMA 7 (2011) 002; arXiv:1010.5709.

\bibitem{S21} S. Sakovich, On two aspects of the Painlev\'{e} analysis, Int. J. Analysis 2013 (2013) 172813; arXiv:solv-int/9909027.

\bibitem{S22} S. Sakovich, Integrability study of a four-dimensional eighth-order nonlinear wave equation, arXiv:1607.08408.

\bibitem{HO} H. Harada, S. Oishi, A new approach to completely integrable partial differential equations by means of the singularity analysis, J. Phys. Soc. Jpn. 54 (1985) 51--56.

\bibitem{K} A. Karasu-Kalkanl\i, Painlev\'{e} classification of coupled Korteweg--de~Vries systems, J. Math. Phys. 38 (1997) 3616--3622.

\bibitem{S23} S.Yu. Sakovich, Coupled KdV equations of Hirota--Satsuma type, J. Nonlinear Math. Phys. 6 (1999) 255--262; arXiv:solv-int/9901005.

\bibitem{S24} S.Yu. Sakovich, Addendum to: Coupled KdV equations of Hirota--Satsuma type, J. Nonlinear Math. Phys. 8 (2001) 311--312; arXiv:nlin/0104072.

\bibitem{KKS1} A. Karasu-Kalkanl\i, A. Karasu, S.Yu. Sakovich, A strange recursion operator for a new integrable system of coupled Korteweg--de~Vries equations, Acta Appl. Math. 83 (2004) 85--94; arXiv:nlin/0203036.

\bibitem{S25} S. Sakovich, A note on the Painlev\'{e} property of coupled KdV equations, Int. J. Part. Diff. Eqns. 2014 (2014) 125821; arXiv:nlin/0402004.

\bibitem{ST1} S.Yu. Sakovich, T. Tsuchida, Symmetrically coupled higher-order nonlinear Schr\"{o}dinger equations: singularity analysis and integrability, J. Phys. A 33 (2000) 7217--7226; arXiv:nlin/0006004.

\bibitem{ST2} S.Yu. Sakovich, T. Tsuchida, Coupled higher-order nonlinear Schr\"{o}dinger equations: a new integrable case via the singularity analysis, arXiv: nlin/0002023.

\bibitem{ST3} S.Yu. Sakovich, T. Tsuchida, A new integrable system of symmetrically coupled derivative nonlinear Schr\"{o}dinger equations via the singularity analysis, arXiv:nlin/0004025.

\bibitem{KKS2} A. Karasu-Kalkanl\i, A. Karasu, S.Yu. Sakovich, Integrability of a generalized Ito system: the Painlev\'{e} test, J. Phys. Soc. Jpn. 70 (2001) 1165--1166; arXiv:nlin/0102030.

\bibitem{KKSST} A. Karasu-Kalkanl\i, A. Karasu, A. Sakovich, S. Sakovich, R. Turhan, A new integrable generalization of the Korteweg--de~Vries equation, J. Math. Phys. 49 (2008) 073516; arXiv:0708.3247.

\bibitem{X} G.Q. Xu, The integrability for a generalized seventh-order KdV equation: Painlev\'{e} property, soliton solutions, Lax pairs and conservation laws, Phys. Scr. 89 (2014) 125201.

\bibitem{S26} S.Yu. Sakovich, On conservation laws and zero-curvature representations of the Liouville equation, J. Phys. A 27 (1994) L125--L129.

\bibitem{HNW} A.N.W. Hone, V. Novikov, J.P. Wang, Generalizations of the short pulse equation, arXiv:1612.02481.

\end{thebibliography}
\end{document}